# Time-lapse Very Long Baseline Interferometry Imaging of the Close Active Binary HR 1099

Walter W. Golay[1,2], Robert L. Mutel[1], and Evan E. Abbuhl[1]
[1] Department of Physics & Astronomy, University of Iowa, 30 N. Dubuque St., Iowa City, IA 52242, USA; wgolay@cfa.harvard.edu
[2] Center for Astrophysics | Harvard & Smithsonian, 60 Garden St., Cambridge, MA 02138, USA

## Abstract

We report multiepoch astrometric very long baseline interferometry observations of the chromospherically active binary HR 1099 (V711 Tau, HD 22468) at six epochs over 63 days using the Very Long Baseline Array at 22.2 GHz. We determined hourly radio positions at each epoch with a positional uncertainty significantly smaller than the component separation. The aggregate radio positions at all epochs define an ellipse in the comoving reference frame with an inclination $i = 39^\circ.5^{+3.6}_{-3.5}$ and longitude of ascending node $\Omega = 212^\circ \pm 22^\circ$. The ellipse center is offset from the Third Gaia Celestial Reference Frame position by $\Delta\alpha = -0.81^{+0.25}_{-0.37}$, $\Delta\delta = 0.45^{+0.23}_{-0.25}$ mas. All radio centroids are well displaced from the binary center of mass at all epochs, ruling out emission from the interbinary region. We examined the motion of the radio centroids within each epoch by comparing hourly positions over several hours. The measured speeds were not statistically significant for five of the six epochs, with $2\sigma$ upper limits in the range 200–1000 km s$^{-1}$. However, for one flaring epoch, there was a $3\sigma$ detection $v_\perp = 228 \pm 85$ km s$^{-1}$. This speed is comparable to the mean speed of observed coronal mass ejections on the Sun.

*Unified Astronomy Thesaurus concepts:* Stellar coronae (305); Interferometric binary stars (806); Magnetic stars (995); Plasma astrophysics (1261); Radio astrometry (1337); Radio continuum emission (1340); RS Canum Venaticorum variable stars (1416); Spectroscopic binary stars (1557); Starlight polarization (1571); Starspots (1572); Stellar magnetic fields (1610)

*Supporting material:* animation

## 1. Introduction

Chromospherically active binaries (CABs) are close binary star systems with at least one late-type spectral component that shows signs of enhanced magnetic activity (e.g., Umana et al. 1998; Strassmeier 2001; Güdel 2002, 2009; Benz & Güdel 2010). Ultraviolet, X-ray, and radio luminosities are typically several orders of magnitude greater than solar values and are highly variable. The most common explanation for enhanced activity in these systems is orbital to spin angular momentum transfer driving a powerful internal dynamo via rapid rotation (Schrijver & Zwaan 1991; Moss & Tuominen 1997; Moss et al. 2002; Moss 2005). Magnetic reconnection provides the mechanism to accelerate electrons to high energies, heating chromospheric and coronal plasma with consequent cooling via thermal and nonthermal radiation at all frequencies.

Nonthermal radio emission from active stars is a unique probe of the highest-energy (∼MeV) particle populations in the coronal magnetoactive plasma. The physical properties and extent of the emitting plasma can be inferred both by modeling the observed spectra (e.g., Falla et al. 1994; Jones et al. 1994; Trigilio et al. 2001; Golay et al. 2023) and by directly measuring the spatial distribution in the binary system using very long baseline interferometry (VLBI; e.g., Mutel et al. 1985; Massi et al. 1988; Lestrade et al. 1993; Lebach et al. 1999, 2012; Ransom et al. 2002, 2005; Peterson et al. 2010, 2011; Abbuhl et al. 2015; Climent et al. 2020). The dominant radiation mechanism at centimeter wavelengths is often interpreted as gyrosynchrotron emission from mildly relativistic electrons filling a region comparable in size to the active star radius, with a modest (∼10–100 G) magnetic field (Dulk 1985; Hewitt & Melrose 1986; Mutel et al. 1998; Güdel 2002).

RS CVn binaries are among the most active CABs (Popper & Ulrich 1977; Hall 1978). They consist of an evolved G- or K-class subgiant tidally locked to a main-sequence companion, with orbital periods of several days to a few weeks. The subgiant typically has cool photospheric spots, enhanced chromospheric emission, and strong, complex magnetic fields (Donati et al. 1992; Donati 1999). However, this canonical description has significant uncertainty: the hot plasma might be dominant on the early-type companion (Drake et al. 2014), since nonaxisymmetric dynamos are possible in solar-like stars (Viviani et al. 2018), or in the interbinary region (Stawikowski & Glebocki 1994; Graffagnino et al. 1995; Trigilio et al. 2001).

HR 1099 (V711 Tau, HD 22468) is among the closest (29.4 pc; Gaia Collaboration et al. 2021) and most well-studied RS CVn binaries. The system is composed of a K1 IV and G5 IV–V (Bopp & Fekel 1976; Fekel 1983) binary pair that is tidally locked in a 2.837 day orbit (Mayor & Mazeh 1987). Photometric observations of HR 1099 show that the face of the active K1 IV star is covered by many optical spots that migrate poleward toward a persistent large dark region (Hatzes et al. 1996; Strassmeier & Bartus 2000; Petit et al. 2004b; Berdyugina & Henry 2007; Kozhevnikova & Alekseev 2015). The complex surface magnetic field of the primary has been mapped using Zeeman–Doppler imaging (Donati et al. 1992; Vogt & Hatzes 1996; Donati 1999; Vogt et al. 1999), uncovering the presence of large, axisymmetric regions where

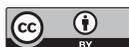

Table 1
Fixed Parameters

| Parameter | Symbol | Value | Uncertainty[a] | References |
|---|---|---|---|---|
| Spectral type | ⋯ | K1 IV + G5 IV–V[b] | ⋯ | (1) |
| Primary radius | $r_1$ | 3.74 $R_\odot$ (0.59 mas) | 0.08 $R_\odot$ (0.01 mas) | (2) |
| Secondary radius | $r_2$ | 1.14 $R_\odot$ (0.18 mas) | 0.08 $R_\odot$ (0.01 mas) | (2) |
| Primary mass | $m_1 \sin^3 i$ | 0.2256 $M_\odot$ | 0.0016 $M_\odot$ | (3) |
| Secondary mass | $m_2 \sin^3 i$ | 0.1752 $M_\odot$ | 0.0011 $M_\odot$ | (3) |
| Primary semimajor axis | $a_1 \sin i$ | $1.8915 \times 10^6$ km (0.0126 au) | $0.0050 \times 10^6$ km | (3) |
| Secondary semimajor axis | $a_2 \sin i$ | $2.4335 \times 10^6$ km (0.0163 au) | $0.0078 \times 10^6$ km | (3) |
| Eccentricity | $e$ | 0.0 | ⋯ | (3) |
| Reference epoch[c] | $T_{\rm ephem}$ | HJD = 2457729.7084 | 0.0017 | (3) |
| Orbital period | $P$ | 2.837711 days | 0.000066 day | (3) |
| R.A. proper motion | $\mu_\alpha \cos\delta$ | −32.2464 mas yr$^{-1}$ | 0.036 mas yr$^{-1}$ | (3) |
| decl. proper motion | $\mu_\delta$ | −162.0732 mas yr$^{-1}$ | 0.032 mas yr$^{-1}$ | (4) |
| Parallax | $\Pi$ | 33.9783 mas | 0.0349 mas | (4) |

**Notes.**
[a] We list uncertainties for convenience. Given their small amplitudes relative to the measured values, we take these parameters to be fixed throughout the analysis presented in this work.
[b] The secondary falls between luminosity classes IV and V.
[c] Time when the K1 IV star is at the ascending node, defined as the phase origin.

**References.** (1) Fekel (1983); (2) Donati (1999); (3) Strassmeier et al. (2020); (4) Gaia Collaboration et al. (2021).

the magnetic field is mainly azimuthal, suggesting that dynamo processes may be distributed throughout the whole convective zone (Petit et al. 2004b).

Maximum entropy maps of photospheric light curves show that starspots occur on both stars, although on the G5 IV–V star, they are less pronounced (García-Alvarez et al. 2003). X-ray observations show periodic flux variation that peaks when the K1 IV component is in front (Audard et al. 2001). A number of X-ray and multiwavelength studies have suggested that observed flares are associated with stellar activity (e.g., Ayres & Linsky 1982; Jones et al. 1996; Robinson et al. 1996; Ayres et al. 2001; García-Alvarez et al. 2003; Osten et al. 2004; Perdelwitz et al. 2018).

HR 1099 was one of the first binary stars ever detected at radio wavelengths (Owen et al. 1976; Feldman et al. 1978). The quiescent radio spectrum has a peak near 10 GHz, a power-law negative spectral index above the peak, and mild circular polarization ($V/I \sim 0.2$). The spectrum is well fit with a power-law gyrosynchrotron emission model consisting of mildly relativistic electrons in a magnetic field of ∼250 G located in a magnetospheric volume about half that of the K1 IV star (Golay et al. 2023). Strong radio flares modify the observed spectrum (Umana et al. 1995; Richards et al. 2003), flattening the spectral index and decreasing the fractional circular polarization, consistent with emission from plasma with higher-energy electrons in a smaller emitting volume (Mutel et al. 1998; García-Sánchez et al. 2003). At lower frequencies, highly circularly polarized flares have been detected (White & Franciosini 1995; Slee et al. 2008; Pritchard et al. 2021). These flares, which have been observed from many other RS CVn binaries (Toet et al. 2021), are consistent with coherent emission from an electron-cyclotron maser mechanism similar to planetary auroral radio emission (Treumann 2006), although a plasma mechanism cannot be ruled out (Toet et al. 2021).

There is increasing interest in stellar coronal mass ejections (CMEs), since they would play a critical role in assessing the habitability of exoplanets orbiting their parent stars (Moschou et al. 2019). However, given that we have currently only detected solar CMEs, it is unclear what role enhanced activity, binarity, or evolutionary status plays in the spectral signatures of stellar CMEs. Solar CMEs (Bastian et al. 2001) have shown cospatial and coeval X-ray and radio emissions (Gary et al. 2018), but this may not apply to active binaries. However, some recent works are starting to provide evidence for stellar CMEs. Moschou et al. (2017) interpreted shadowing of a large X-ray flare as indirect detection of a stellar CME, while Inoue et al. (2023) measured prominence eruption speeds from the CAB V1355 Orionis (using Hα line shifts) that were much larger than the escape speed, suggestive of a CME event.

Highly temporally resolved VLBI maps of the radio emission from active stars with sufficiently high spatial resolution may provide the first direct radio observation of a stellar CME, including their speeds and direction. Using a global array of radio antennas at centimeter wavelengths, contemporary phase-referenced VLBI observations (Diamond 1995; Fomalont & Kogan 2005) can determine the structure and positions of radio sources with an angular resolution and astrometric accuracy of order 100 μas or less (Pradel et al. 2006; Reid & Honma 2014; Reid 2022). This results in a positional accuracy significantly smaller than the binary separation for stars whose distance is less than ∼100 pc. However, to place a radio map on the same coordinate grid as the binary, the astrometric parameters (proper motion, parallax, fiducial position at a reference epoch) of the binary must be known with comparable accuracy.

The recently released Gaia DR3 catalog (Gaia Collaboration et al. 2021) solved this problem by providing astrometric parameters with uncertainties comparable to the positional uncertainty of phase-referenced radio maps ($\lesssim$100 μas; see Table 1). This allows accurate registration of the stellar positions on the radio maps, subject to a small correction to account for the orientation and spin of the Gaia Celestial Reference Frame (GCRF) with respect to the radio-based International Coordinate Reference Frame (ICRF; Lindegren 2020; Chen et al. 2023).

Although HR 1099 was observed several times with VLBI arrays in the 1980s and 1990s (e.g., Lestrade et al. 1984; Mutel et al. 1984, 1985; Massi et al. 1988; Lestrade et al. 1993), high-

**Table 2**
HR 1099 Best-fitting Hourly Offsets, Angular Sizes, Flux Densities, and Fractional Polarization

| Epoch | R.A. (GCRF3)[a] (Mid-epoch) | Decl. (GCRF3)[a] (Mid-epoch) | JD 2400000.5+ | $\Delta\alpha$[b] (mas) | $\Delta\delta$[b] (mas) | Phase | FWHM[c] ($R_K$) | Flux[d] (mJy) | $V/I$[e] |
|---|---|---|---|---|---|---|---|---|---|
| A 2013 Nov 13 | $03^h36^m47{.}^s26061$ | $00°35'13{.}''6960$ | 56609.189 | $+1.68 \pm 0.37$ | $-0.66 \pm 0.57$ | 0.36 | 1.4 | 28.5 | 0.22 |
| | | | 56609.234 | $+1.20 \pm 0.18$ | $-0.87 \pm 0.43$ | | | | |
| | | | 56609.281 | $+1.18 \pm 0.52$ | $-0.96 \pm 0.33$ | | | | |
| | | | 56609.327 | $+1.73 \pm 0.45$ | $-0.57 \pm 0.20$ | | | | |
| B 2013 Nov 19 | $03^h36^m47{.}^s26058$ | $00°35'13{.}''6934$ | 56615.204 | $+0.45 \pm 0.16$ | $-1.63 \pm 0.27$ | 0.45 | 0.9 | 53.6 | 0.20 |
| | | | 56615.251 | $+0.62 \pm 0.24$ | $-1.12 \pm 0.41$ | | | | |
| | | | 56615.297 | $-0.02 \pm 0.18$ | $-1.16 \pm 0.28$ | | | | |
| C 2013 Nov 26 | $03^h36^m47{.}^s26054$ | $00°35'13{.}''6903$ | 56622.221 | $+1.21 \pm 0.07$ | $+0.71 \pm 0.15$ | 0.91 | 1.0 | 164.8 | 0.03 |
| | | | 56622.267 | $+1.15 \pm 0.05$ | $+0.41 \pm 0.10$ | | | | |
| | | | 56622.313 | $+0.96 \pm 0.07$ | $+0.54 \pm 0.11$ | | | | |
| D 2013 Dec 31 | $03^h36^m47{.}^s26033$ | $00°35'13{.}''6744$ | 56658.150 | $-0.03 \pm 0.59$ | $-0.96 \pm 0.26$ | 0.57 | 0.5 | 22.0 | 0.22 |
| | | | 56658.193 | $+0.03 \pm 0.30$ | $-0.71 \pm 0.41$ | | | | |
| | | | 56658.239 | $+0.25 \pm 0.46$ | $-0.85 \pm 0.16$ | | | | |
| E 2014 Jan 1 | $03^h36^m47{.}^s26032$ | $00°35'13{.}''6739$ | 56659.100 | $+0.53 \pm 0.34$ | $+0.42 \pm 0.63$ | 0.92 | 0.9 | 18.2 | 0.20 |
| | | | 56659.146 | $+0.30 \pm 0.12$ | $+0.12 \pm 0.58$ | | | | |
| | | | 56659.192 | $+0.64 \pm 0.07$ | $+0.26 \pm 0.15$ | | | | |
| F 2014 Jan 16 | $03^h36^m47{.}^s26023$ | $00°35'13{.}''6673$ | 56674.084 | $+1.66 \pm 0.21$ | $-0.16 \pm 0.28$ | 0.19 | 1.3 | 39.6 | 0.22 |
| | | | 56674.130 | $+1.60 \pm 0.13$ | $-0.18 \pm 0.26$ | | | | |
| | | | 56674.177 | $+1.65 \pm 0.14$ | $-0.34 \pm 0.20$ | | | | |

**Notes.**
[a] The binary center-of-mass position at the epoch's midpoint computed using the reference position given in Table 3 and the proper motion and parallax from Table 1.
[b] Offset of radio centroid with respect to the mid-epoch GCRF3 position (second and third columns).
[c] FWHM angular size of a circular Gaussian model fit to the calibrated visibilities in units of K1 IV star radius.
[d] We determine fluxes relative to our primary calibrator CTA 26, for which we assume a flux density of 2.5 Jy by extrapolating the light curve from Angelakis et al. (2019).
[e] Fractional circular polarization (ratio of Stokes parameters $V/I$).

angular-resolution radio maps were not published until almost 20 yr later. Ransom et al. (2002) observed HR 1099 at 8.4 GHz using the Very Long Baseline Array (VLBA; and four additional telescopes) at two epochs, once during the decay stage of a flare and once when the source was quiescent. The flaring epoch map had two peaks separated by the binary component separation. They interpreted this as two sources that are either in the corona of the K1 IV star straddling either side of the disk or emission from separate regions on each star. Using snapshot 2–4 hr time-sliced images, they found that the two peaks rotated counterclockwise, which they speculated was due to the corotation of the radio-emitting region with the binary system.

In this work, we report multiepoch VLBI phase-referenced observations of the HR 1099 binary. We determine the hourly locations of the radio emission within the comoving frame of the binary at each epoch to determine both the position relative to the binary components and possible motion in the corotating frame. Our observational scheme is outlined in Section 2. We detail the astrometric VLBI calibration and reduction to positions in Section 3. In Section 4, we describe our method of determining the missing orbital parameters, constraining source motion, and characterizing the polarization properties, and we discuss our results. In Section 5, we compare our astrometric position determination (referenced to the Third ICRF, ICRF3, radio frame of reference) with that of the Third GCRF (GCRF3), compare our results with previous VLBI studies of HR 1099, and discuss the analysis of radio centroid hourly motion in the context of solar and stellar CMEs, and we conclude in Section 6.

We include extended discussions of our astrometric centroid fitting procedure in Appendix A, corrections to atmospheric phase delays in Appendix B, and our orbital fitting routine in Appendix C. We provide a GitHub repository[3] and an archival Zenodo[4] listing of the code used to generate the figures in this work. Individual figure captions include links denoted by ✎.

## 2. Observations

HR 1099 was observed at six 10 hr epochs with the National Radio Astronomy Observatory's (NRAO) VLBA. We scheduled the observations to sample the orbital phase at approximately equal spacings so that any source variation from changes in the binary phase, e.g., a source corotating with the binary, could be observed. The receivers were centered at 22.2 GHz and recorded using the polyphase filter bank (PFB). The PFB records in 16 dual-polarization 32 MHz subchannels with 2-bit sampling for an aggregate data rate of 2048 Mb s$^{-1}$. Table 2 contains the Julian Date at the midpoint of each observing epoch, along with the fiducial positions predicted by the Gaia proper motions and parallax at each epoch.

We used the nodding phase-referencing scheme (Lestrade et al. 1990), scheduling observations that cycled between the primary phase calibrator (CTA 26) and the target HR 1099 using a 3 minute cycle consisting of 0.5 minute observing the

---
[3] https://wwgolay.github.io/HR1099-timelapse-vlbi/README.html
[4] doi:10.5281/zenodo.10395762

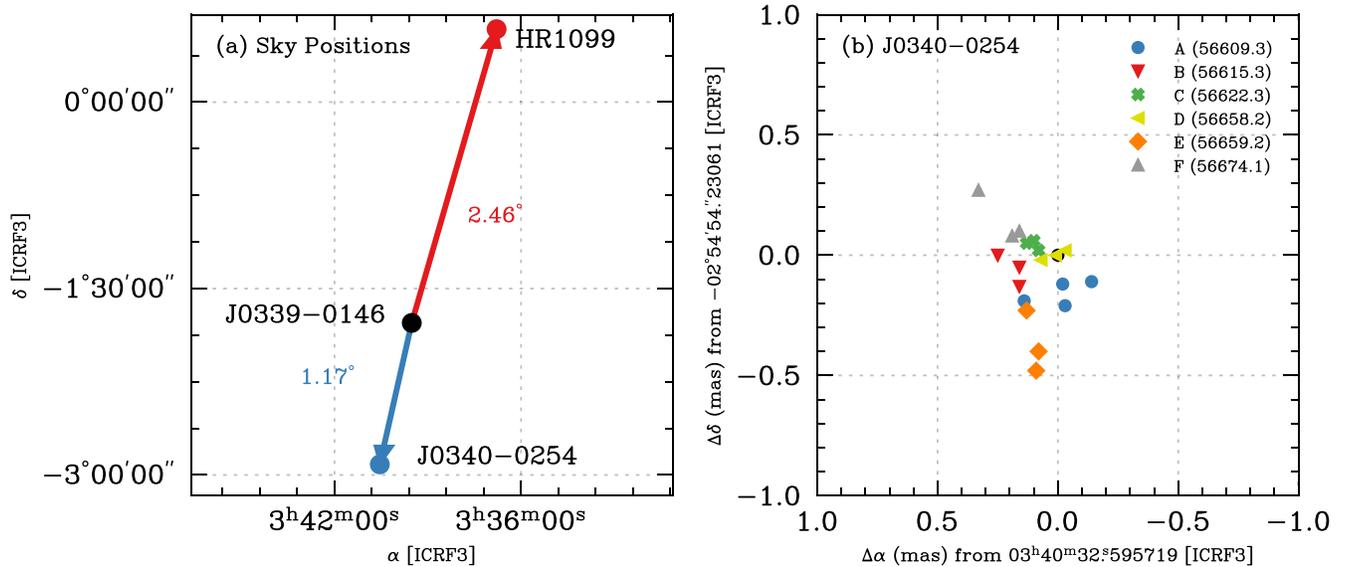

**Figure 1.** Left: the relative positions and angular separations between the primary phase calibrator J0339–0146 (CTA 26), HR 1099 (V711 Tau), and the secondary calibrator J0340–0254. Right: hourly positions of J0340–0254 relative to its ICRF3 position at each epoch. Uncorrected tropospheric delays likely cause these shifts. Each hourly position shift of J0340–0254 was multiplied by −2.1 in each coordinate and applied to the fitted position of HR 1099 (see text for details).

**Table 3**
Source Coordinates

| Source | Frame | $\alpha$ | $\sigma$ (mas) | $\delta$ | $\sigma$ (mas) | Epoch | References |
| --- | --- | --- | --- | --- | --- | --- | --- |
| J033930.9−014635 (CTA 26) | ICRF3 | $03^h39^m30\overset{s}{.}937787$ | 0.03 | $-01°46'35\overset{''}{.}80420$ | 0.03 | 2000.0 | (1) |
| J034032.5−025454 (J0340−0254) | ICRF3 | $03^h40^m32\overset{s}{.}595719$ | 0.07 | $-02°54'54\overset{''}{.}23061$ | 0.20 | 2000.0 | (1) |
| HR 1099 (V711 Tau) | GCRF3 | $03^h36^m47\overset{s}{.}256031$ | 0.03 | $+00°35'13\overset{''}{.}35052$ | 0.03 | 2016.0 | (2) |

**References.** (1) Charlot et al. (2020); (2) Gaia Collaboration et al. (2021).

calibrator followed by 2.5 minutes observing HR 1099. Additionally, once every 15 minutes, a nearby secondary calibrator (J0340−0254) was observed for 2 minutes to check the stability of the phase-referencing scheme. The primary calibrator is separated by 2°.46 from HR 1099 and by 1°.17 from the secondary calibrator J0340−0254. The relative alignment of the three sources is nearly a straight line, with the primary calibrator located close to the line connecting HR 1099 and J0340−0254 (Figure 1(a)). Source coordinates are listed in Table 3.

### 3. Data Analysis

#### 3.1. Editing and Visibility Calibration

We calibrated the observations using the NRAO Astronomical Image Processing Software ($\mathcal{AIPS}$; Greisen 2003). First, the data were correlated with a fixed reference position at all epochs, since precise astrometric parameters were not available at the time of the observations. Following correlation, we corrected the phase center for proper motion and parallax using the newly available precise astrometric values from Gaia DR3 (Gaia Collaboration et al. 2021) using the $\mathcal{AIPS}$ task CLCOR. This ensured that the phase center at all epochs was fixed in the comoving frame of HR 1099's center of mass.

Next, we applied several data quality flags. We flagged all visibilities observed at low elevation (<30°), since astrometric uncertainty rapidly increases at high air mass (Reid 2022). Next, we flagged all visibilities on projected baselines for which HR 1099 was highly resolved (ranging from a maximum of 150 to 250 M$\lambda$, depending on the epoch), and consequently, the visibility phases were very noisy. Finally, we flagged all baselines containing the VLBA station Saint Croix, since the visibility phases were highly unstable, probably due to a large wet troposphere component (see Pradel et al. 2006).

We then applied standard VLBI amplitude and delay-rate corrections. We first applied a priori amplitude and digital sampling corrections based on the system temperatures and NRAO-supplied gain curves. Next, Earth pole orientation variations were corrected using cataloged Earth orientation parameters (IERS & Dick 2020). We then applied parallactic angle corrections and manual phase calibration to determine instrumental delays and rates. Finally, global fringe-fitting solutions were performed at each scan of the primary phase calibrator to find the variable complex gain corrections. These corrections were smoothed and applied to the target and secondary phase calibrator (see Diamond 1995). We then exported the data for inspection and further processing.

After editing and full calibration, we used the difference mapping program difmap (Shepherd 1997) to create self-calibrated maps for both HR 1099 and the phase calibrator sources. These maps were used to model the visibility phases, as explained below.

#### 3.2. Astrometric Calibration

A primary goal of these observations is to determine the radio emission's precise location in the binary system's reference frame at each epoch. In addition, we wish to test for the possible motion of the radio centroid within each epoch, as might occur if the radio emission is associated with a stellar analog to solar CME events. To do this, we partitioned each epoch's visibilities into hourly time slices, keeping only those hours not flagged by the abovementioned flagging criteria. This resulted in either three or four hourly data sets per epoch.

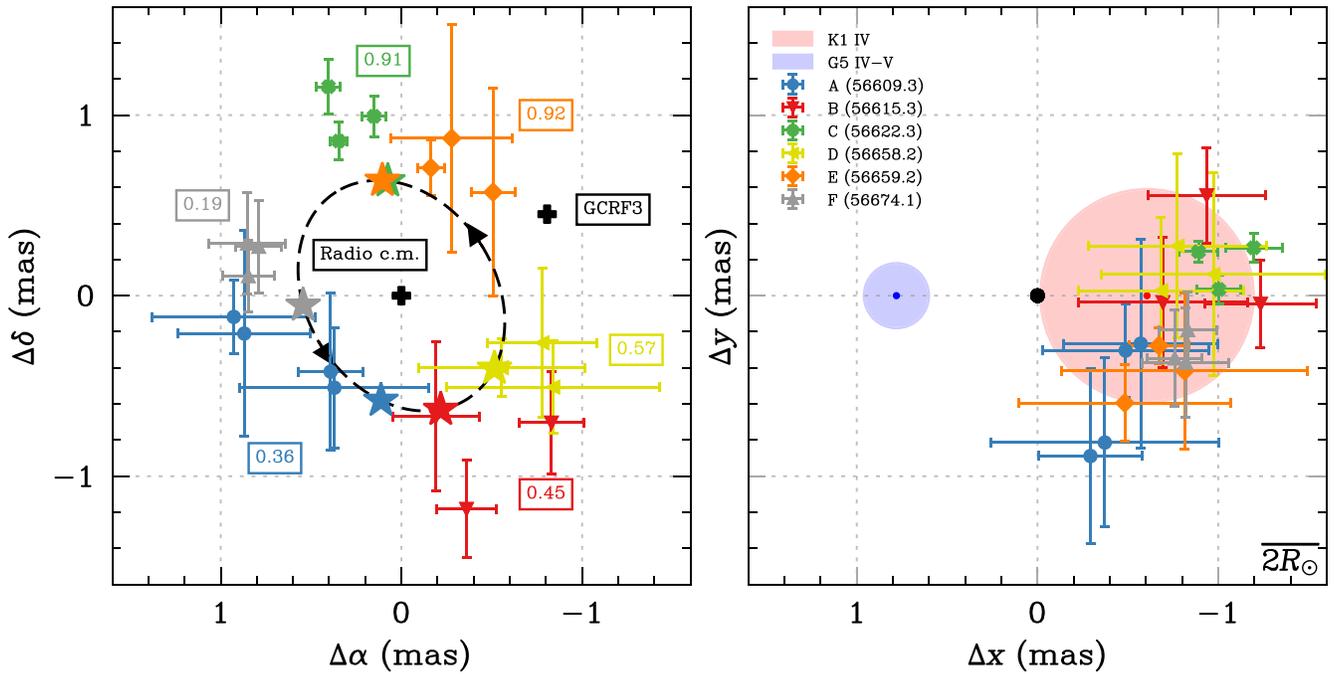

**Figure 2.** Left: hourly radio centroid offset positions at all epochs with respect to the Gaia DR3 coordinates at that epoch. The dashed line is the K1 IV orbit using the fitted orbital parameters in Table 4, with arrows indicating orbital motion direction. The radio centroid location, location of the primary (K1 IV) star center, and orbital phase are color-coded by epoch: blue (A), red (B), green (C), yellow (D), orange (E), and gray (F). The black crosses labeled Radio c.m. and GCRF3 are the comoving positions of the center of the fitted orbit and the Gaia DR3 position, respectively. Right: the same radio positions displayed in the corotating frame of the binary, along with the epoch-averaged location of the primary (red circle), secondary (blue circle), and center of mass (black dot) in the corotating frame.

We calculated the difference between the observed visibility phase at each sampled spatial frequency $(u, v)$ and the corresponding phase computed from a model brightness for each hourly time slice. The model comprises the self-calibrated radio map for that epoch, which is Fourier transformed to determine the visibility phase at each observed $(u, v)$ coordinate. The model objective function to be minimized is the square-summed difference between the observed and model phases weighted by the observed phase uncertainties. The model fit minimization scheme uses a "brute-force" search, evaluating the objective function on a square grid of model offsets with 0.01 mas spacing centered at the phase center. The best-fit offsets correspond to the coordinates of the cell with the smallest value.

Determining the uncertainty in these offsets is nontrivial. Generally, we do not expect the two-dimensional best-fitting centroid location probability distribution to be Gaussian. We use the probability maps constructed from the brute-force grid search to account for this effect and measure the likelihood function for each independently determined position. For an extended discussion of the fitting procedure, see Appendix A.

### 3.2.1. Correction for Tropospheric Effects

When applying the above procedure to the secondary calibrator J0340–0254, we find nonnegligible offsets (~0.1–0.5 mas) from the ICRF3 coordinates for this source, as seen in Figure 1(b). We attribute these shifts to random variations in the tropospheric path length that are not accounted for in the VLBA correlator model (Martí-Vidal et al. 2010; Reid 2022). If we assume a tropospheric isoplanatic patch extending over several degrees (Monnier & Allen 2013, Section 7.2.4.1 and Table 7.2) and that the lines of sight to both calibrator sources and HR 1099 intercept this patch, we can use J0340–0254's observed angular offsets to estimate the residual tropospheric correction on the line of sight to HR 1099.

Figure 1(a) displays the relative positions of the calibrators and target. The angular separation of HR 1099 is 2.1 times greater than and in the opposite direction of J0340–0254. Hence, for each hourly time slice, we apply a shift to HR 1099's computed offset using the observed shift of J0340–0254 at the same hour but multiplied by the factor −2.1. Appendix B describes this correction in greater detail. HR 1099's hourly position offsets after this correction are listed in Table 2.

## 4. Results

### 4.1. Orbit Orientation and Inclination

In the comoving frame of the binary, the radio centroid position offsets clearly delineate an elliptical trajectory with motion in a counterclockwise rotation, as seen in Figure 2 (left). From optical spectroscopy, several orbital elements of the binary system are known with high precision, e.g., the semimajor axis, orbital period, and eccentricity (Table 1). The orbital inclination is less well determined, with published values ranging from $33° \pm 2°$ (Fekel 1983) to $40° \pm 5°$ (Donati 1999), while the longitude of the ascending node ($\Omega$) has remained completely undetermined, since the angular extent of the binary is too small to resolve optically.

For an inclined circular orbit, $\Omega$ equals the angle between the meridional line to the north celestial pole and the direction to the ascending node as measured from the ellipse's center. The intersection of the line of nodes on the projected orbit has two locations, either of which could be the location of the ascending node. Hence, the location of the ascending node is $\pm\pi$ degenerate. We resolve this degeneracy by assuming that the radio emission is associated with the K1 IV subgiant rather than the G5 IV–V secondary, a plausible assumption given many previously published studies showing that strong magnetic fields are dominantly on the K1 IV star (e.g., Donati et al. 1992; Vogt & Hatzes 1996; Donati 1999; Vogt et al. 1999; Petit et al. 2004a) and hence are likely the site of the radio emission. Another supporting argument is that for the two other CABs that have been previously mapped using astrometric VLBI, Algol (Lestrade et al. 1993) and IM Pegasi (Ransom et al. 2012), the radio emission was shown to be centered on the subgiant.

#### 4.1.1. Orbit Model Fitting Algorithm

We fit the hourly radio centroid positions with an inclined, circular orbit by varying four parameters: sky orientation ($\Omega$), inclination ($i$), and the net offset ($\Delta\alpha$, $\Delta\delta$) between the radio-determined orbit center (ICRF3 frame) and the optical position given by the GCRF3. The fitting scheme used an objective function given by the weighted summed squared differences between the VLBI positions and those calculated using fixed orbital parameters (period, semimajor axes, mass ratios) and variable parameters (orbit orientation ($\Omega$), inclination angle, and offsets between the ICRF3 and GCRF3 frames). Note that the radio offsets are in the comoving frame of the binary. See Appendix C for further discussion of our likelihood function selection that is agnostic to associating the radio emission with either star.

Since the system's inclination is somewhat uncertain (Section 4.1), we express this uncertainty with a Gaussian prior centered on 40° with $\sigma = 5°$. Additionally, to account for a net offset between the GCRF3 and the ICRF3, the model includes two parameters representing a shift in R.A. $\Delta\alpha$ and decl. $\Delta\delta$. Both parameters have a prior that is uniform over the centroid position-fitting grid-search region defined in Section 3.2.

We sample the posterior using the Python implementation of the Monte Carlo Markov Chain (MCMC) emcee (Foreman-Mackey et al. 2013) with the "stretch-step" algorithm and the default step-length probability distribution. We initialize a small (1% of parameter values) hyper-Gaussian ball of 100 walkers centered on an estimated set of parameters $i$, $\Omega$, $\Delta\alpha$, and $\Delta\delta$. The walkers take 1000 discarded "burn-in" steps and 4000 "production" steps. Solution convergence is determined by inspecting the autocorrelation time and the acceptance fraction as a function of step number, where a robust solution was defined as a chain length exceeding at least $50\times$ the autocorrelation time in all four parameters. In all cases, 4000 steps were sufficient to achieve convergence. Table 4 lists best-fit solutions for all four fitted parameters. The orbital inclination is very close to the value previously reported by Donati (1999) based on optical spectroscopy. The radio-optical frame offsets are discussed in Section 5.1.

### 4.2. Source Locations in the Binary Corotating Frame

Using the orbital parameters, we can transform the on-sky radio centroid offsets to the corotating frame of the binary, as shown in the right panel of Figure 2. The centroids are all located on or near the K1 IV star disk and not in the interbinary region, which argues against models that posit emission in

**Table 4**
Parameters Derived from Best-fit Orbit

| Parameter | | Value | Uncertainty |
|---|---|---|---|
| Radio-opt. R.A. offset | $\Delta\alpha$ | −0.81 mas | +0.25, −0.37 |
| Radio-opt. decl. offset | $\Delta\delta$ | 0.45 mas | +0.23, −0.25 |
| Orbital inclination | $i$ | 39°.5 | +3°.6, −3°.5 |
| Long. ascending node[a] | $\Omega$ | 212° | ±22° |

**Note.**
[a] The measured value has a 180° ambiguity. The listed angle assumes that the radio emission is associated with the K1 IV subgiant. See Section 4.1 for an explanation.

close active binaries driven by joint magnetospheres (e.g., Graffagnino et al. 1995; Siarkowski et al. 1996; Richards et al. 2012; Hill 2017; Singh & Pandey 2022).

### 4.3. Component Motion

Since each epoch has multiple hourly positions, we can explore whether there is statistically significant motion in the frame of the K1 IV star within each epoch. To investigate this possibility, we rotate and center the hourly probability maps for each hourly radio position in the corotated frame. We use a linear speed model parameterized by an initial position ($x_0$, $y_0$) and a linear velocity vector ($v_\parallel$, $v_\perp$) along axes defined parallel and perpendicular to the line connecting the binary components.

To fit this model, we use a likelihood function that is defined by the product of the evaluated value of the probability map for each position. We transform the position maps into the corotated frame, and then, for a given intercept and velocity, we predict the position at each observed time. The likelihood for that time is the probability map's value at the predicted position. Finally, we multiply this likelihood with the likelihoods from the other positions in that epoch. The intercepts have a prior that is uniform over the centroid position-fitting grid-search region defined in Section 3.2. The joint posterior probability distributions for the velocities are computed using the same MCMC procedure described in Section 4.1.

These distributions are shown in Figure 3, where the black lines in each panel indicate the null hypothesis (no motion). For five of the six epochs, the posterior distributions are consistent with the null hypothesis, i.e., no statistically significant projected motion on either axis. However, for epoch C (middle left panel of Figure 3), the perpendicular speed ($v_\perp = 228^{+84}_{-86}$ km s$^{-1}$) lies on the 99% contour and hence likely exhibits statistically significant motion. This is noteworthy since this epoch is also the only one exhibiting a radio flare (the flux density is 3–4× the value at any other epoch; Table 2).

### 4.4. Radio Lobe Sizes

We measured a characteristic radio emission angular size at each epoch by fitting a one-dimensional Gaussian model to the visibility amplitude versus the $uv$ baseline length, averaged over all projected angles (Table 2). The mean sizes were similar at all epochs, ranging from 0.5× to 1.4× the K1 IV star radius. The source morphology (as reconstructed by self-calibration during the model generation of the fitting procedure from Section 3.2) is displayed for each epoch in Figure 4. (An animated version of this plot is available on GitHub[5] and in the

---
[5] https://wwgolay.github.io/HR1099-timelapse-vlbi/notebooks/4_radio_maps.html#generate-movie

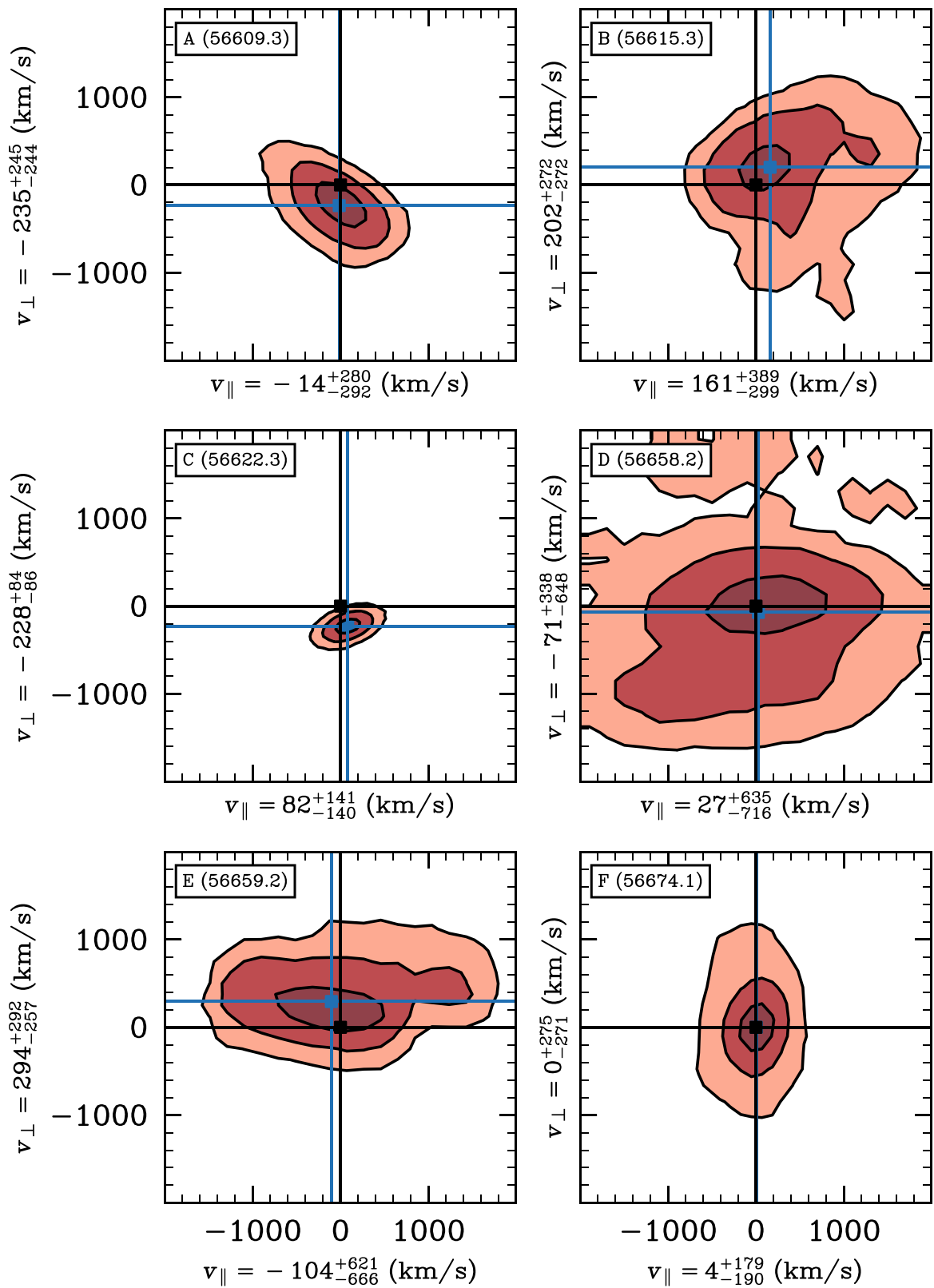

**Figure 3.** Joint posterior distributions of R.A. and decl. velocity components in the corotating reference frame. Contours are displayed at 39%, 87%, and 99%. The blue line shows the sample medians, and the black line is the null hypothesis (no motion). The axis labels are at 16%, 50%, and 84%. At all epochs but C, measured speeds along each axis are consistent with the null hypothesis of no statistically significant projected motion in either axis. However, at epoch C, ($v_\perp = 228^{+84}_{-86}$ km s$^{-1}$) lies on the 99% contour, suggesting statistically significant motion.

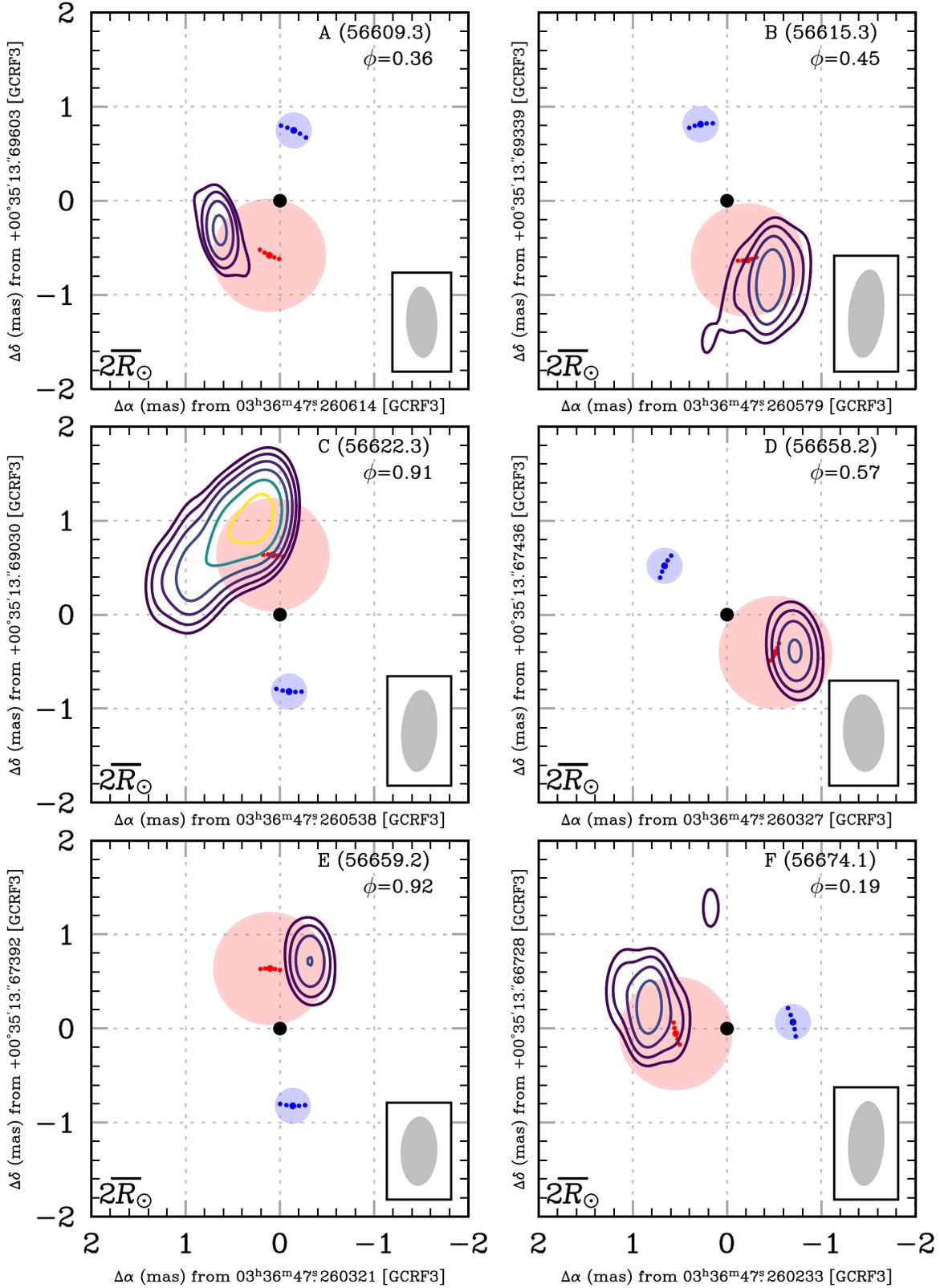

**Figure 4.** Self-calibrated CLEAN maps of the HR 1099 radio emission. The contours are plotted over each component of the binary's expected position given the fitted orbital parameters and radio-optical offsets in Table 4, given a counterclockwise orbit. The origin is the binary's position predicted by the Gaia DR3 parameters (Table 3). The colored dots indicate the corresponding star's position at 1 hr intervals. Note that there is a 180° degeneracy in Ω, which amounts to an association with one binary component or the other (see Section 4.1). Contours are color-scaled and shown at 4%, 8%, 16%, 32%, and 64% of the peak flux over all epochs (fluxes are listed in Table 2). An animated version of this plot is available at GitHub (https://wwgolay.github.io/HR1099-timelapse-vlbi/notebooks/4_radio_maps.html#generate-movie). The animation is 48 s in length and displays a continuous orbit of the binary stars, briefly slowing down during the phase range for which our observations coincide, which is represented by each panel in this figure.

(An animation of this figure is available.)

online material. The emitting region is resolved in all six epochs.)

The median angular size, 0.9 times the K1 IV star radius, agrees very well with the angular size $2.5\,R_\odot = 0.75\,R_K$ derived by Golay et al. (2023) using a power-law gyrosynchrotron radio emission model evaluated at 22 GHz.

### 4.5. Location of Right Circular Polarization and Left Circular Polarization Emission Regions

We performed an analysis similar to Peterson et al. (2010) to determine if two displaced, oppositely circularly polarized radio-emitting regions exist in HR 1099. We begin by individually self-calibrating the Stokes $I$, right circular-only (RR), and left circular-only (LL) data using the difmap script automap and saving the outputs separately for each epoch. We then fit the RR and LL self-calibrated models to the RR and LL components of the self-calibrated Stokes $I$ data using the same procedure outlined in Section 3.2. The resulting probability maps represent a measure of the net offset of that polarization's emission region from the mean position of the total emission region regardless of polarization. Finally, the resulting RR and LL probability maps were convolved to generate the likelihood of separate polarized positions. When fitting to the opposite circular polarization data, we find no statistically significant evidence of a net separation ($\Delta\theta < 0.1$ mas) between the centroids of the RR and LL polarizations for all six epochs.

## 5. Discussion

### 5.1. Radio-optical Position Offset

The GCRF3 (Gaia Collaboration et al. 2021) is defined by the positions of a large number of extragalactic sources (mostly quasars) that define a kinematically nonrotating frame. The radio counterpart is the ICRF3 (Ma et al. 1998; Charlot et al. 2020), whose positions are determined by astrometric VLBI observations. The mutual alignment of these catalogs at the submilliarcsecond level has proven challenging for several reasons, including the frequency-dependent brightness structure of the quasar cores (Liu et al. 2021) and color and brightness biases. Most quasars are bluer than stars and faint (99% have $G$ magnitudes > 17 mag), and they have a very different distribution on the sky than stars. These differences in magnitude, color, and sky distribution will likely produce small shifts of the image centroids, which could propagate into systematic position errors (Liu et al. 2021).

High-precision phase-referenced VLBI observations of stars can provide an independent astrometric data set that avoids the biases inherent in quasar position measurements (Lindegren 2020; Chen et al. 2023; Makarov et al. 2023). Our observations of HR 1099 span a time range of 65 days, which is too short to measure parallax or proper motion precisely. However, position measurements over a well-sampled range of orbital phases were sufficient to determine the size and orientation of the orbit on the sky (Figure 2).

Assuming that the radio emission is located on the K1 IV primary (see Section 4.1), the radio-derived orbit center is the binary's barycenter, which is equal to the center of the K1 IV star orbit. This position (in the comoving frame) can be compared with the Gaia DR3 catalog position, which averages the binary's brightness centroid over many Gaia observations. Since the brightness of the K1 IV star is about 1 mag brighter than the G5 IV–V star (Donati et al. 1992), there is a small positional bias (about 0.1 mas) toward the K1 IV star, but since the Gaia sampling cadence is random with respect to the orbital period, this bias is negligible when averaged over many samples and hence orientations. Hence, we ignore this effect and take the Gaia catalog position as the best estimate of the binary barycenter in the optical (GCRF3) frame.

The resulting offsets ($\Delta\alpha = -0.81^{+0.25}_{-0.37}$, $\Delta\delta = 0.45^{+0.23}_{-0.25}$ mas) are consistent with the coordinate transformation from ICRF3 to GCRF3 as given by Lindegren (2020, Equations (6) and (7)) using the parameters of solution B in Table 7 of Chen et al. (2023). The predicted offsets ($-0.52$, $+0.58$ mas) are within one standard deviation of our measured values, supporting their solution.

### 5.2. Comparison with Previous VLBI Maps

Comparing the results of Ransom et al. (2002) with our maps is challenging because the source structure may be frequency-dependent, so the double-lobed map at 8 GHz during a flaring epoch is not necessarily in disagreement with the single-component images we observed at all epochs, including the flare at epoch C that we observed at 22 GHz. However, both the counterclockwise (CCW) rotation of their western component and the vector direction (approximately northerly between orbital phases 0.67 and 0.76) agree with our orbit solution (Figure 2). If their western component is identified with the K1 IV primary, the eastern component will be located at or near the G5 IV–V secondary, supporting the suggestion of Ransom et al. (2002) that radio emission, at least during very large flares, may originate in a shared magnetosphere containing the radio-emitting volume.

By contrast, our corotated maps (Figure 2) clearly indicate that the radio emission centroids are located near the center or on the far side of the K1 IV star (Section 4.1) and not in the interbinary region. This is the case even at epoch C, which exhibited a large flare but also had the most complex brightness structure (Figure 4). The brightness distribution is not dual-lobed, as seen in the Ransom et al. (2002) flare-epoch map, but there is a clear extension pointing southeast, although not in the direction or distance of the secondary star.

Whether enhanced radio emission (and other enhanced activity indicators) arises from an interbinary region in CABs is still an open question. While most radio astrometric studies have concluded that the emission originates at the subgiant (Lestrade et al. 1993; Ransom et al. 2012; Abbuhl et al. 2015), several studies (e.g., Graffagnino et al. 1995; Siarkowski et al. 1996) argue that X-ray emission could arise in the interbinary region. Unfortunately, since there is only a modest correlation between X-ray and radio flares (at least for HR 1099; Osten et al. 2004), inferring the location of the radio emission from X-ray data is problematic.

One intriguing conjecture is that quiescent radio emission originates in regions of strong magnetic flux on the K1 IV primary star, but intense flares are generated at both stars, possibly resulting from magnetic field entanglement between the primary and secondary.

### 5.3. Component Motion: Comparison with Solar CMEs

CME events on the Sun are highly correlated with solar flares, with more energetic X-ray flares corresponding to faster and more massive CMEs (Moschou et al. 2019). However, although X-ray flares are also commonly detected on other active stars, the Sun has been the only star to allow direct CME observations. Since CMEs are a potential threat to the habitability of exoplanets orbiting active stars, there is strong interest in detecting and characterizing stellar CMEs (e.g., Kay et al. 2016).

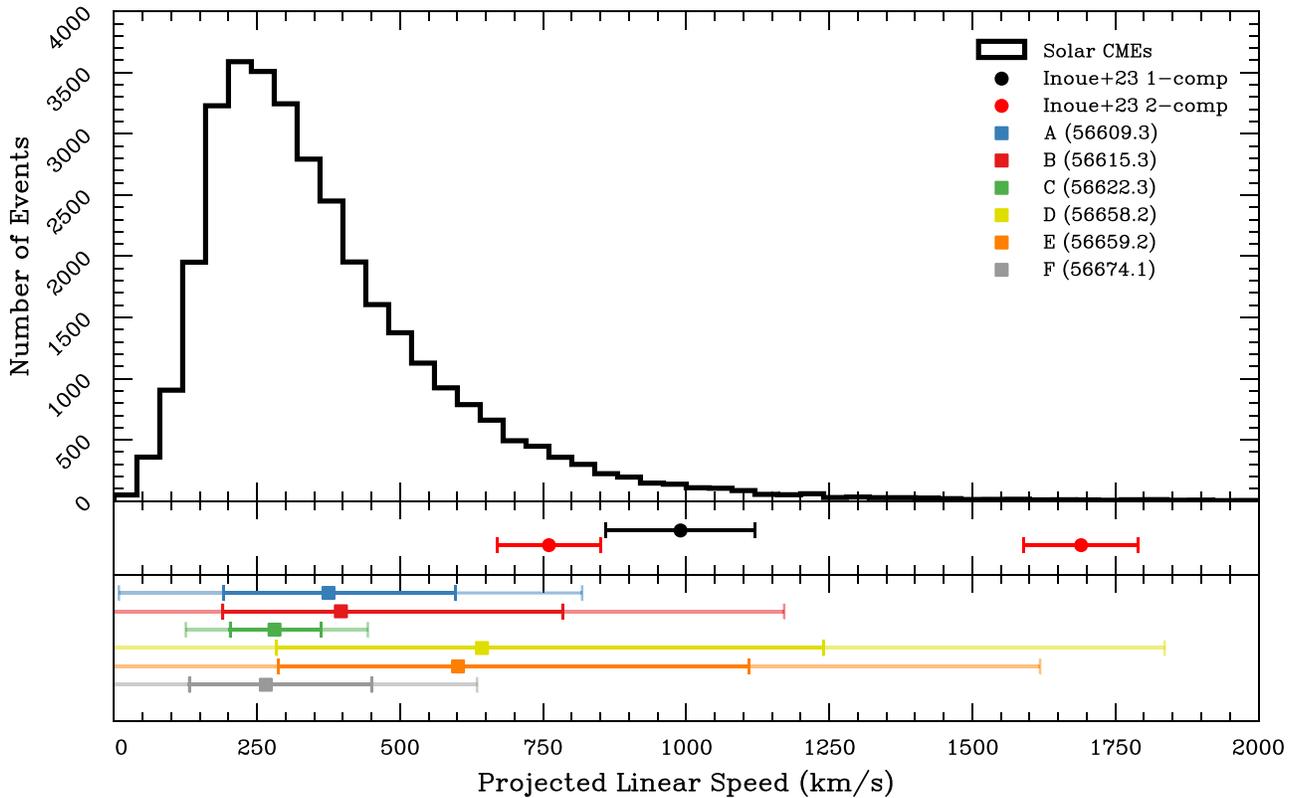

**Figure 5.** Top: comparison between the measured solar CME linear speed distribution from the SOHO LASCO CME catalog (Gopalswamy et al. 2009; omitting flares marked as "poor" or "very poor"). Middle: the reported velocities of a CME associated with a superflare on another RS CVn–type star, V1355 Orionis (Inoue et al. 2023). Bottom: the 16th, 50th, and 84th percentiles of the velocity distribution magnitudes (speeds) from Figure 3. We include a semitransparent extension of each percentile by its magnitude to illustrate the significance of the measured motion. Five of the six epochs support the null hypothesis that no statistically significant motion exists. There is a 3σ detection of a nonzero speed during the flare in epoch C.

Our measurement of a shift in the radio centroid position during a flaring event (epoch C), if interpreted as the motion of a CME structure, invites comparison with solar CME speeds. Unfortunately, the solar analog is not obviously comparable, since the system's binarity and later-type subgiant are not easily related to the Sun. However, although flare emission from the Sun is much less luminous, coronal loops on the Sun may have similar plasma properties to CAB coronae, including similar Alfvén speeds (Kansabanik 2023). With this proviso, comparing our projected velocity measurements with eruptive solar events is instructive.

There have been several studies of the distribution of observed solar CME speeds. Yurchyshyn et al. (2005) analyzed 5 yr of the Large Angle and Spectrometric Coronagraph Experiment on the Solar and Heliospheric Observatory (LASCO SOHO) entailing white-light data of 4315 CMEs. They found that both accelerating and decelerating CMEs are reasonably well described by a lognormal speed distribution peaking at ~430 km s$^{-1}$. Other analyses of LASCO SOHO flares showed correlations between the flare speeds, peak flux, and fluence properties and the soft X-ray emission (Salas-Matamoros & Klein 2015).

Studies focusing on the kinematics of centimeter-wave type IV radio emission on the Sun are less statistically robust. Equivalent emission processes to the gyrosynchrotron radiation observed from CABs peak at a much lower frequency on the Sun. Nonetheless, several observations of the continuum centimeter-wave companion to white-light CMEs have been made. Bastian et al. (2001) compared a 1998 April 20 flare at optical and radio wavelengths and found that the radio and optical CME loops were similar in position, morphology, and expansion speed but were not the same. Analyses of a radio/optical CME from 2001 April 15 show a similar trend and emphasize that electron acceleration occurs at low altitudes (Maia et al. 2007; Démoulin et al. 2012).

Figure 5 shows the hourly radio centroid motions at all epochs compared with the probability distribution of solar CME events (Gopalswamy et al. 2009) and inferred speed of two Hα flares on the RS CVn binary V1355 Ori (Inoue et al. 2023). All measured speeds along both axes are consistent with no motion at the 2σ confidence level except for the lone flaring epoch (C), which is similar to the mean solar CME speed.

## 6. Summary

We have presented a six-epoch astrometric phase-referenced VLBI study of the RS CVn binary HR 1099. We determined the hourly positions of the radio-emitting region with submilliarcsecond accuracy over a well-sampled range of binary orbital phases. We summarize our results as follows.

1. The radio centroids clearly define an ellipse, with the radio centroid moving counterclockwise. By fitting the observed positions with an inclined circular orbit, we find a best-fit inclination angle $i = 39\overset{\circ}{.}5^{+3.6}_{-3.5}$ and longitude of ascending node $\Omega = 212° \pm 22°$ assuming the K1 IV subgiant (rather than the G5 IV–V secondary) is the locus of the radio emission.
2. Since the radio centroid is displaced from the center of mass by at least 0.5 mas during all epochs, it is not located in the interbinary region at any epoch, and no physically motivated coordinate transformation could place it in the interbinary region.
3. After correction for proper motion and parallax, the offset between the ICRF3 (Charlot et al. 2020) radio centroid of HR 1099 and the Gaia DR3 (GCRF3; Gaia Collaboration et al. 2021) position at each epoch is $\Delta\alpha = -0.81^{+0.25}_{-0.37}$, $\Delta\delta = 0.45^{+0.23}_{-0.25}$ mas. This is in good agreement with ICRF3–GCRF3 offsets computed using the recently reported frame orientation parameters of Chen et al. (2023).
4. We tested for the motion of the radio centroids by comparing hourly positions at each epoch. For five of the six epochs, the measured speeds both along and normal to the orbital plane were not statistically significant, with $2\sigma$ upper limits in the range 200–1000 km s$^{-1}$. However, for one flaring epoch, there was a $\sim3\sigma$ detection $v_\perp = 228 \pm 85$ km s$^{-1}$. This speed is comparable to the mean speed of CMEs observed on the Sun (Gopalswamy et al. 2009) and inferred H$\alpha$ flare velocities on another RS CVn binary, V1355 Orionis (Inoue et al. 2023).
5. The radio centroid positions for left and right circularly polarized emission coincide within 0.1 mas at all epochs, likely inconsistent with models in which oppositely circularly polarized emission originates at the feet of large-scale coronal loops.

## Acknowledgments

We thank Antonino Lanza for a helpful discussion on the longitudinal distribution of nonaxisymmetric magnetic fields and Paul Cristofari for contextualizing the capabilities of Zeeman–Doppler imaging. We thank the anonymous referee for the helpful feedback. The National Radio Astronomy Observatory (https://public.nrao.edu/) is operated by Associated Universities, Inc., under cooperative agreement with the National Science Foundation. This work made use of the DiFX software correlator developed at Swinburne University of Technology as part of the Australian Major National Research Facilities program (Deller et al. 2011). This work has made use of data from the European Space Agency (ESA) mission Gaia (https://www.cosmos.esa.int/gaia), processed by the Gaia Data Processing and Analysis Consortium (DPAC; https://www.cosmos.esa.int/web/gaia/dpac/consortium). Funding for the DPAC has been provided by national institutions, in particular the institutions participating in the Gaia Multilateral Agreement. This CME catalog is generated and maintained at the CDAW Data Center by NASA and the Catholic University of America in cooperation with the Naval Research Laboratory. SOHO is a project of international cooperation between ESA and NASA. This research has made use of NASA's Astrophysics Data System (https://ui.adsabs.harvard.edu/).

## Data Availability

The uncalibrated visibilities are available for download from the NRAO Data Archive (https://data.nrao.edu/portal) under project code BM392 (PI: R. Mutel). We provide Table 2 as our observed best-fitting radio centroids for each epoch and MJD. A GitHub repository (https://github.com/WWGolay/HR1099-timelapse-vlbi) and a user-friendly website (https://wwgolay.github.io/HR1099-timelapse-vlbi) that reproduces the results presented here are available. An archival release of the code coincident with publication is also hosted at Zenodo (see footnote 4). The position-fitting code is available upon request from the corresponding author.

*Facility:* Very Long Baseline Array (VLBA).

*Software:* $\mathcal{AIPS}$ (version 31DEC2023; Greisen 2003), Astropy (v6.0.0; Astropy Collaboration et al. 2013, 2018, 2022), Cmcrameri (v1.7; Crameri 2023), Corner.py (v2.2.2; Foreman-Mackey 2016), `difmap` (v2.5q; Shepherd 1997), Emcee (v3.1.4; Foreman-Mackey et al. 2013), ghp-import (v2.1.0; Davis 2022), Jupyter Book (v0.15.1; Executable Books Community 2020), Matplotlib (v3.8.2; Hunter 2007), Numpy (v1.26.2; Harris et al. 2020), Scipy (v1.11.4; Virtanen et al. 2020), Skyfield (v1.24; Rhodes 2023), and Smplotlib (v0.0.9; Li 2023).

## Appendix A
## Centroid Position Fitting

Reconstructing the best possible source model with a sparsely and nonuniformly sampled *uv* plane presents a challenge for interferometric imaging. Observations of sources that are static in morphology and position can generally be well reconstructed as the Earth's rotation fills in the *uv* plane over time when combined with standard imaging techniques, e.g., `CLEAN` (Högbom 1974) and the maximum entropy method (Cornwell & Evans 1985). However, sources with a resolvable change in their position over the course of an observation complicate the process of constructing an accurate source model. It is difficult to distinguish between changes in the visibility phase and amplitudes that are due to differences in the source morphology that become resolved as the Fourier plane is filled in by the Earth's rotation versus changes caused by a source in motion.

Most astronomical objects do not have a resolvable change in the source position over the maximum possible observation length, even on the longest baselines. However, sources with angular velocities that are of order one synthesized beam per observation length must be carefully imaged. High-proper-motion sources and close binary systems can easily exceed the relevant angular velocity. If a source's proper motion is well defined, it can be corrected in imaging. However, in the case of CABs where not all the orbital parameters are known, correcting for this effect is possible but challenging. Motion that would not be corrected by translating to the binary's comoving frame, such as orbital corotating motion, can be at sufficiently high velocities to be resolved on timescales as short as an hour. Since HR 1099 is a CAB with a short period (see Table 1), the source model must be allowed to change as a function of time.

We fit directly to the visibility data to allow for a dynamic model. The source model is composed of the sum of an arbitrary number of elliptical Gaussians with widths and rotation angles defined by the synthesized beam for that observation. A complex visibility at some spatial frequency $\mathcal{V}(u, v)$ is the Fourier transform of a positive-definite real source function on the sky $I_\nu(\alpha, \delta)$. An elliptical Gaussian source function with width parameters $\sigma_\alpha$ and $\sigma_\delta$, and rotation angle $\phi$ is given by

$$I_\nu(\alpha, \delta; \sigma_\alpha, \sigma_\delta, \phi) = \frac{I_{\nu,0}}{\sigma_\alpha \sigma_\delta} \exp\{-\pi \boldsymbol{x}'^2\},$$

where $\boldsymbol{x}' = \mathcal{U}\mathcal{R}_\phi \boldsymbol{x} \equiv \begin{pmatrix} 1/\sigma_\alpha & 0 \\ 0 & 1/\sigma_\delta \end{pmatrix} \begin{pmatrix} \cos\phi & \sin\phi \\ -\sin\phi & \cos\phi \end{pmatrix} \begin{pmatrix} \alpha \\ \delta \end{pmatrix}.$ (A1)

It can be shown for an elliptical Gaussian that

$$\mathcal{V}(u, v) = I_{\nu,0} \int_{-\infty}^{\infty} \exp[-\pi \alpha'^2 - 2\pi i \alpha' u'] d\alpha'$$
$$\times \int_{-\infty}^{\infty} \exp[-\pi \delta'^2 - 2\pi i \delta' v'] d\delta',$$
where $\begin{pmatrix} u' \\ v' \end{pmatrix} \equiv \mathcal{U}\mathcal{R}_\phi \begin{pmatrix} u \\ v \end{pmatrix}.$ (A2)

By factoring $u'$ and $v'$ out of the exponentials and noting that the remaining integrals become the integral of a normalized Gaussian, we find that

$$\mathcal{V}(u, v) = I_{\nu,0} e^{-\pi(u'^2 + v'^2)}.$$ (A3)

The elliptical Gaussian components can be located at any position. The shift theorem allows us to compute the visibilities for an arbitrarily placed elliptical Gaussian. We are interested in the Fourier transform of a model defined by an arbitrary number $N$ of elliptical Gaussian components. Since the Fourier transform is a linear operator, we can simply sum over these components:

$$\mathcal{V}_{\text{tot}}(u, v) = \sum_{n=1}^{N} I_{\nu,0,n} e^{-\pi(u_n'^2 + v_n'^2)} \times e^{2\pi i(\Delta\alpha_n u + \Delta\delta_n v)}.$$ (A4)

Note that the multicomponent approach improves over the CASA task uvmodelfit (CASA Team et al. 2022), which can only accept a single-component model.

We use the output of difmap's self-calibration script automap to define the model for a given source and epoch. automap returns a list of fluxes paired with offsets from the phase center (called CLEAN components), which we use to define the amplitudes and positions of the Gaussian components. Since automap does not return individual widths and angles for each component, we use the synthesized beam parameters for that epoch and source as the widths and angles for all CLEAN components. The complete model is the sum of all components, a method similar to the Gaussian kernel density estimate. Note that automap operates on a completely unflagged data set except for the flags from $\mathcal{AIPS}$ and the Kitt Peak VLBA station for epochs D, E, and F due to a misconfiguration in the cross-hand readouts during the observation.[6]

To reduce computation time, we make two simplifications to the model fitting. The source model only uses the components from automap up to a cutoff. The cutoff is where the cumulative flux converges to the final normalized flux. Additionally, we downsample from the 2 s correlator dump time to 60 s by taking the vector average of the visibilities as the new value and the visibility amplitude standard deviation as the uncertainty for each record.

To find the best-fitting centroid, we take a "brute-force" approach of defining a square grid of R.A. and decl. offsets from the phase center ($\Delta\alpha, \Delta\delta$), and we evaluate the goodness of fit of the source model to the observed visibilities at each point. For any elliptical Gaussian centered at the origin, all visibility phases will be zero, since the complex vector has no imaginary component. This means the visibility phases contain only astrometric information. We define the likelihood function as the reduced sum of squares of the differences in phase between the model and data (simple $\chi_\nu^2$), normalized by the uncertainties from downsampling to 60 s records. We record the likelihood function's value at each point in the grid to generate a probability surface map that we save as a function that linearly interpolates between the evaluated points. We inspect the "dirty maps" of J0340–0254 and HR 1099 to identify each source's approximate offset from the phase center to determine the optimal grid-search parameters. We use a grid extent of ±1 mas from the phase center for J0340–0254 and ±2 mas for HR 1099 and a grid step size of 0.01 mas for both sources.

We use a two-step method to determine if there is apparent source motion during an observation. First, we fit for the centroids on a per-epoch, per-source basis. Then, we divide each epoch and source into individual ∼10–15 minute subobservation scan cycles. We then group these scans into sets of five, which will be used to fit for a single position. Given the observing strategy outlined in Section 2, this is approximately 1 hr of data. We then fit positions for each grouped scan set and selected a contiguous set of scan sets in the center of the observation that best agreed with the per-epoch position. Note that the model is only epoch-dependent and not regenerated for each scan set. Both approaches were attempted and resulted in negligible differences.

## Appendix B
## Correcting Atmospheric Phase Delays

VLBI observations are susceptible to different atmospheric phase delays above each antenna. The ionospheric phase delay approximately scales with the inverse square of the frequency. Because the ionosphere is dispersive, multifrequency observations can be used to model and correct this effect. At higher frequencies ($\gtrsim$5 GHz), the ionospheric phase delays become negligible relative to other effects. The wet and dry components of the troposphere dominate the phase delays above this frequency range. Most phase delays are from the dry component made up of atmospheric gases, whose effect can be predicted using the local temperature and pressure and will only vary by ∼1% in a few hours. The water vapor content that defines the wet component depends on the local weather conditions and causes delays that are, on average, smaller than the dry component. However, the wet component varies much more randomly and rapidly, making correcting its effect challenging (Leick 2003).

The VLBA sites do not have water vapor radiometers, so the precise wet troposphere properties are unknown. The VLBA correlator applies a seasonal model to correct for these phase delays. Astrometric observations are typically limited by the uncertainty and nonlocality of this model (Reid 2022). The astrometric error scales approximately linearly with the angular separation within an individual tropospheric wedge (approximately a few degrees on the sky; Pradel et al. 2006). $\mathcal{AIPS}$ provides two tasks for mitigating these phase delays using two different strategies. DELZN uses observations of multiple calibrators to compute a slowly varying tropospheric phase delay over the entire sky ($\mathcal{AIPS}$ Memo #110: Mioduszewski & Kogan 2009). However, given the significant overhead of observing eight or more sources every 4 hr (∼1 additional hr), we chose not to pursue this approach in our experimental design. The $\mathcal{AIPS}$ task ATMCA attempts to locally correct tropospheric delays in the region of the sky around the target. By observing multiple calibrators straddling the target in various potential scan configurations (Section 2 in $\mathcal{AIPS}$ Memo #111: Fomalont & Kogan 2005), one can construct a two-dimensional model for the tropospheric phase gradients in the sky and interpolate them to the target position.

---

[6] See Issue #506 from the VLBA data archive issue page (https://www.vla.nrao.edu/astro/archive/issues/#506).

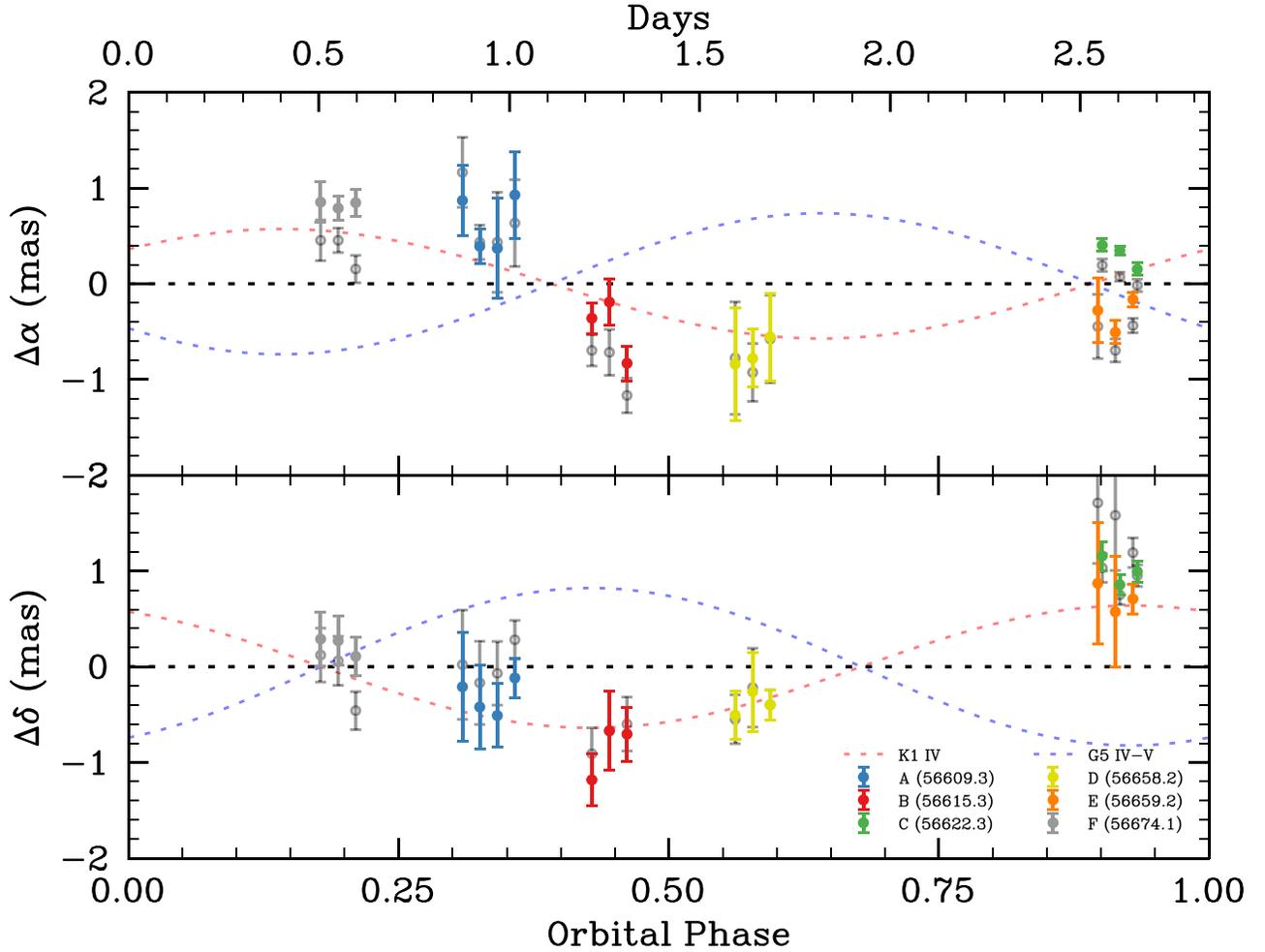

**Figure 6.** Top: the R.A. offsets in the binary's comoving reference frame of our measured radio centroids before (transparent) and after (opaque, colored, and enumerated in Table 2) applying our atmospheric phase delay correction procedure. We plot against the binary's phase to demonstrate that the applied shifts are minor compared to each star's offset (red and blue dashed lines) from the binary center (black dashed line). Bottom: the same plot for the decl. offsets.

Since our observations used standard phase referencing with only a single secondary calibrator (corresponding to case (f) in Section 2 of Fomalont & Kogan 2005), and HR 1099 does not fall in between CTA 26 and J0340−0254, our geometry is not optimal but technically valid for ATMCA, requiring extrapolation of the phase delay rather than interpolation. Because of this geometry and that HR 1099 would not have a high enough signal-to-noise ratio in only a few minutes of integration required to determine phase delays, we applied a similar approach to the ATMCA correction algorithm. We model the subinterval best-fit positions of J0340−0254 as independent measurements of the mean effect of the wet troposphere. We correct this by extrapolating the position offsets to the location of HR 1099 using the angular separation and direction relative to CTA 26. This can be expressed as

$$\Delta \phi_{\mathrm{T}} = \Delta \phi_{\mathrm{C}} \cdot \mathrm{proj}_{\hat{n}_{\mathrm{P} \to \mathrm{T}}} \left( \frac{\boldsymbol{n}_{\mathrm{P} \to \mathrm{T}}}{\boldsymbol{n}_{\mathrm{P} \to \mathrm{S}}} \right), \quad (\mathrm{B1})$$

where $\boldsymbol{n}_{\mathrm{P} \to \mathrm{S}}$ and $\boldsymbol{n}_{\mathrm{P} \to \mathrm{S}}$ are the vectors along the primary calibrator to the secondary calibrator S and the target T and $\hat{n}$ is the unit vector. For HR 1099, this astrometric error factor evaluates to ≃−2.1 given the separations of J0340−0254, HR 1099, and CTA 26. We apply this correction to the HR 1099 positions for each solved position. For each scan set, we take the relative offset of J0340−0254 during the same scan set from its phase center, multiply by the astrometric correction factor of −2.1, and then shift the HR 1099 position by this value. The J0340−0254 positions used to compute these corrections are shown in Figure 1.

Since the final positions are the difference of the HR 1099 positions and the J0340−0254 offsets scaled by the astrometric error factor of −2.1, the final HR 1099 position maps are the convolution of the original HR 1099 map and the J0340–0254 map scaled by the astrometric error factor (Equation B1). We save these maps to be used for fitting to determine HR 1099's orbital parameters (Section 4). The positions and errors reported in Table 2 and displayed in Figure 2 are from two-dimensional Gaussian fits to the final HR 1099 map. However, only the maps and not the values are used in the analysis. In practice, these corrections are small relative to the binary orbit, as shown in Figure 6.

## Appendix C
## Orbital Fitting Routine

Recalling the discussion from Section 4.1, we fit for the longitude of ascending node ($\Omega$), inclination ($i$), and the net offset between the ICRF3 and GCRF3 ($\Delta\alpha$, $\Delta\delta$) under the assumption that the radio emission is associated with the K1 IV star.

To debias the fitting routine from this assumption, we define a likelihood function that is not dependent on the physical properties of either star. This means the likelihood function is agnostic to selections of $\Omega$ that are 180° apart. In general, the semimajor axes of the component stars (and the distance from the binary center of mass at any given time) will not be equal, so the likelihood function must be agnostic to the radial distance of the position. At any given time, each star's position relative to the decl. axis is at an angle $\phi$ and $\phi + \pi$, respectively, so we can find an optimal radially independent model by minimizing the angular difference between the star's position and radio position at each observation time. The inclination $i$ controls the apparent radial distance between the two stars for circular orbits, so it will not be well constrained by this method as slight increases in the semimajor axes are also consistent with lower relative inclination. However, the longitude of ascending node $\Omega$ controls the angle of each star at a given orbital phase, meaning a likelihood function that minimizes the angle is optimized to constraining $\Omega$. As long as the likelihood function's shape over the angle from the center of either star is independent of the position at $\phi$ or $\phi + 180°$, it is also independent of the star selection.

We implement this likelihood function by a coordinate transformation of each position's probability map. The maps are expressed as functions of ($r, \theta$). We then collapse over the radial dimension to give a purely $p(\theta)$ map by taking the maximum value of the probability map along the radial line for each $\theta$. If the likelihood function were to minimize the position relative to the star's angle, it would preferentially select an association of the radio emission with the far side of the star relative to the other binary component. The position angle of one body in a two-body system must be the angle opposite the other body, so a likelihood function that peaks at the position angle would be biased toward the system's far side. Since we do not know where the radio emission will be located on the star's surface or limb, we convolve the $p(\theta)$ map with a top-hat (square) function centered on the star's angle $\phi$ and a 90° total width (45° on either side). This convolution also reduces the preference to minimize any given epoch with more flux (with a better-constrained $p(\theta)$ map) to be near the star's far side.

## ORCID iDs

Walter W. Golay  https://orcid.org/0000-0001-7946-1034
Robert L. Mutel  https://orcid.org/0000-0003-1511-6279